\documentclass[aps,physrev,reprint,superscriptaddress]{revtex4-2}

\usepackage{graphicx}
\usepackage{todonotes}
\usepackage{amsmath}
\usepackage{amsfonts}
\usepackage{xcolor}
\usepackage[colorlinks=true,linkcolor=blue,citecolor=blue,urlcolor=blue]{hyperref}

\begin{document}


\title{Phase-preserving control of Floquet-engineered cavity quantum electrodynamics}


\author{Jia-Wang Yu}
\email{jw.yu@zju.edu.cn}
\affiliation{State Key Laboratory of Silicon and Advanced Semiconductor Materials \textnormal{\&} College of Information Science and Electronic Engineering, Zhejiang University, Hangzhou 310027, China}
\affiliation{ZJU-Hangzhou Global Scientific and Technological Innovation Center, Zhejiang University, Hangzhou 311200, China}
\author{Xiao-Qing Zhou}
\affiliation{Department of Physics, School of Science, Westlake University, Hangzhou 310030, China}
\affiliation{Institute of Natural Sciences, Westlake Institute for Advanced Study, Hangzhou 310024, China}
\author{Zhi-Bo Ni}
\affiliation{State Key Laboratory of Silicon and Advanced Semiconductor Materials \textnormal{\&} College of Information Science and Electronic Engineering, Zhejiang University, Hangzhou 310027, China}
\affiliation{ZJU-Hangzhou Global Scientific and Technological Innovation Center, Zhejiang University, Hangzhou 311200, China}
\author{Xiao-Tian Cheng}
\affiliation{State Key Laboratory of Silicon and Advanced Semiconductor Materials \textnormal{\&} College of Information Science and Electronic Engineering, Zhejiang University, Hangzhou 310027, China}
\affiliation{ZJU-Hangzhou Global Scientific and Technological Innovation Center, Zhejiang University, Hangzhou 311200, China}
\author{Yi Zhao}
\affiliation{State Key Laboratory of Silicon and Advanced Semiconductor Materials \textnormal{\&} College of Information Science and Electronic Engineering, Zhejiang University, Hangzhou 310027, China}
\affiliation{ZJU-Hangzhou Global Scientific and Technological Innovation Center, Zhejiang University, Hangzhou 311200, China}
\author{Hui-Hui Zhu}
\affiliation{State Key Laboratory of Silicon and Advanced Semiconductor Materials \textnormal{\&} College of Information Science and Electronic Engineering, Zhejiang University, Hangzhou 310027, China}
\affiliation{ZJU-Hangzhou Global Scientific and Technological Innovation Center, Zhejiang University, Hangzhou 311200, China}
\affiliation{College of Integrated Circuits, Zhejiang University, Hangzhou 311200, China}
\author{Chen-Hui Li}
\affiliation{State Key Laboratory of Silicon and Advanced Semiconductor Materials \textnormal{\&} College of Information Science and Electronic Engineering, Zhejiang University, Hangzhou 310027, China}
\affiliation{ZJU-Hangzhou Global Scientific and Technological Innovation Center, Zhejiang University, Hangzhou 311200, China}
\affiliation{Research Institute of Intelligent Computing, Zhejiang Lab, Hangzhou 311121, China}
\author{Feng Liu}
\affiliation{State Key Laboratory of Silicon and Advanced Semiconductor Materials \textnormal{\&} College of Information Science and Electronic Engineering, Zhejiang University, Hangzhou 310027, China}
\author{Chao-Yuan Jin}
\email{jincy@zju.edu.cn}
\affiliation{State Key Laboratory of Silicon and Advanced Semiconductor Materials \textnormal{\&} College of Information Science and Electronic Engineering, Zhejiang University, Hangzhou 310027, China}
\affiliation{ZJU-Hangzhou Global Scientific and Technological Innovation Center, Zhejiang University, Hangzhou 311200, China}
\affiliation{College of Integrated Circuits, Zhejiang University, Hangzhou 311200, China}


\date{\today}

\begin{abstract}
We propose a Floquet-engineered framework for the coherent control of the light–matter interaction in a two-level system (TLS) located in a time-modulated cavity. Strictly phase-preserving operation of the TLS-cavity interaction is demonstrated, allowing the interrupt and retrieval of coherent Rabi oscillations without the loss of quantum information. By introducing a phonon reservoir, it is proved that the frequency instability induced from non-Markovian processes does not produce significant phase decoherence during Floquet modulation. Our results provide new insights into the fundamental physics of a driven quantum system and establish Floquet engineering as a powerful tool for coherent quantum information processing.
\end{abstract}



\maketitle

\section{Introduction}
Coherent and tunable control of quantum interactions in the time domain is a central requirement for a wide range of quantum information processing (QIP) tasks, such as quantum state transfer, quantum memory buffering, entanglement distribution, and the realization of scalable quantum networks \cite{kimble2008quantum, aharonovich2016solid, somaschi2016near, flamini2018photonic, gonzalez2024light}. For instance, in cavity quantum electrodynamics (CQED) systems, a single quantum emitter coherently coupled to a local photonic mode acts as a retrievable quantum node, where the emitter-cavity interface enables real-time conversion between flying and static qubits. This capability makes CQED a key building block for high-bandwidth quantum networks based on photonic architectures \cite{kimble2008quantum, reiserer2015cavity, reiserer2022colloquium}.

Realizing the full potential of CQED for high-fidelity QIP requires precise control of both the amplitude and phase of quantum states \cite{liu2023quantum, johne2015control, johne2011single}. In particular, phase preservation is critical in interference-based protocols fundamental to quantum networks, such as Bell state measurements, entanglement swapping, and quantum teleportation \cite{surmacz2006entanglement, khodjasteh2013designing, tiecke2014nanophotonic}. In large-scale quantum networks, even small phase deviations accumulate across gates and memories, becoming a dominant limitation on fidelity. Phase-preserving control is therefore a prerequisite for scalable and high-performance quantum photonic architectures \cite{sun2016quantum}. Despite this critical requirement, existing approaches to controlling interactions in CQED platforms, such as tuning spectral overlap \cite{faraon2010fast, bose2014all, albert2013microcavity, shambat2011ultrafast, jin2014ultrafast} or modulating loss rates \cite{bradford2013spontaneous}, often neglect phase preservation as a primary design constraint. Consequently, these methods frequently induce uncontrolled phases, typically arising from off-resonant evolution after the interaction is switched off. These phase shifts degrade the quantum state fidelity and ultimately limit the scalability and operational flexibility of quantum photonic networks.

Here, we propose a phase-preserving scheme for coherent interaction control between solid-state quantum emitters and cavities based on Floquet engineering \cite{oka2019floquet, rudner2020floquet, yu2026enhancing}. In case emitter or cavity resonances are periodically modulated over time on solid-state platforms \cite{blattmann2014entanglement, buhler2022chip, lukin2020spectrally}, programmable high-fidelity control of light–matter interactions has been widely demonstrated \cite{li2023single, li2024dynamic}. Although time-dependent Hamiltonian inherently generates time-varying phases, the systematic mitigation of these phases during coherent control has not been adequately addressed. We demonstrate that interactions can be switched on and off without imparting uncontrolled phase shifts, enabling seamless interruption and reactivation of coherent processes. This is achieved by employing smoothly varied modulation envelopes that suppress non-adiabatic transitions between instantaneous Floquet modes. Crucially, the periodic phase structure of the Floquet solution, combined with optimized switching control sequence, leads to near-cancellation of the phase contributions accumulated over the total control sequence. As a result, the relative phase between the emitter and the cavity is preserved upon reactivation of the interaction, providing a robust pathway toward realizing programmable quantum photonic networks and high-performance quantum memory architectures.

\section{Model and Theory}

\begin{figure}[h]
    \centering
    \includegraphics[width=0.45\textwidth]{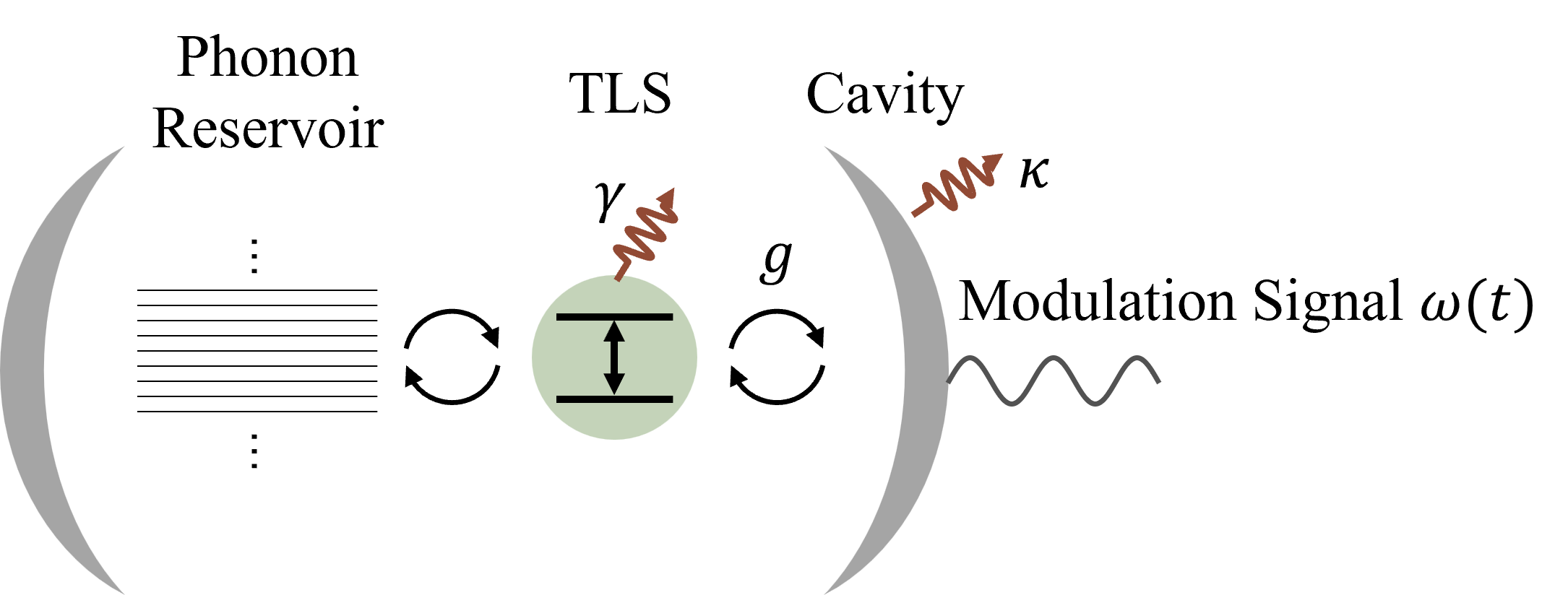}
    \caption{ Schematic of the proposed model with a two-level system (TLS) coupled to a cavity with periodically modulated resonance frequency. A phonon reservoir is included to account for dephasing processes. When the coherent destruction of tunneling (CDT) condition is satisfied, the effective coupling between the TLS and the cavity mode is strongly suppressed.
    }
    \label{fig:model}
\end{figure}

Fig.~\ref{fig:model} schematically illustrates the proposed system, which consists of a two-level system (TLS) coherently coupled to a cavity whose resonance frequency is dynamically modulated in time.

\subsection{Two-Level System and Modulated-Cavity System}

We first consider the coherent interaction between a two-level system (TLS) and a single cavity mode in the absence of phonon coupling. In the absence of modulation, the TLS transition frequency and the cavity resonance are assumed to be on resonance at the optical frequency $\omega_0$. To simplify the description, we work in a rotating frame with respect to the bare optical frequency $\omega_0$ and adopt the rotating-wave approximation (RWA). In this frame, the system Hamiltonian takes the form \cite{yang2004interactions, blattmann2014entanglement, saharyan2023propagating}
\begin{equation}\label{eq:labHs}
    H_S(t) = \omega(t)\, a^\dagger a + g \left(a^\dagger \sigma + \sigma^\dagger a\right),
\end{equation}
where $a^\dagger$ ($a$) is the creation (annihilation) operator of the cavity mode, and $\sigma^\dagger$ ($\sigma$) is the raising (lowering) operator of the TLS. The time-dependent cavity frequency $\omega(t)$ describes a periodic modulation at frequency $\Omega$, and $g$ denotes the TLS-cavity coupling strength. The RWA description is valid in the regime $\omega_0 \gg g$, and provided that the modulation amplitude and the modulation frequency $\Omega$ remain much smaller than $\omega_0$, such that rapidly oscillating counter-rotating terms can be neglected. Under these conditions, the total excitation number operator $N_{\mathrm{tot}} = a^\dagger a + \sigma^\dagger \sigma$ commutes with $H_S(t)$ at all times. Therefore, once the system is initialized in the single-excitation manifold, the dynamics remain confined to this subspace.

To analyze the dynamics under periodic driving, we employ the Floquet formalism~\cite{goldman2014periodically, eckardt2015high, martin2017topological}. The time-periodic Hamiltonian $H_S(t)$ can be decomposed as
\begin{equation}
    H_S(t) = \sum_{n} e^{in\Omega t} H^{[n]}_S .
\end{equation}
Here, the operators $H^{[n]}_S$ act entirely within the physical Hilbert space and describe the usual system degrees of freedom, whereas the phase factors $e^{in\Omega t}$ shift the Fourier sector from arbitrary $m$ to $m+n$, since $e^{in\Omega t} e^{im\Omega t} = e^{i(m+n)\Omega t}$. In the Floquet construction, these two structures are promoted to operators acting on independent Hilbert spaces, forming an extended Hilbert space given by the tensor product of the physical space and an infinite-dimensional Floquet (Fourier) space spanned by the orthonormal basis states $\{|m\rangle\}$, $m\in\mathbb{Z}$. In this representation, the time-independent Floquet Hamiltonian takes the form
\begin{equation}\label{eq:laFbHsF}
    \mathcal{H}_S = I \otimes \Omega N + \sum_{n} H^{[n]}_S \otimes T^n,
\end{equation}
where $T = \sum_{m} |m+1\rangle\langle m|$ is the shift (raising) operator and $N = \sum_{m} m |m\rangle\langle m|$ is the number operator, both acting on the Floquet space, and $I$ denotes the identity operator on the physical Hilbert space. The operator $T^n$ shifts the Floquet index by $n$, with $T^n = \sum_{m} |m+n\rangle \langle m|$, and physically represents transitions between Floquet sectors separated by $n\Omega$. The first term in Eq. (\ref{eq:laFbHsF}), $I \otimes \Omega N$, gives the energy spacing between Floquet sector, while the second term, $\sum_{n} H^{[n]}_S \otimes T^n$, couples different Floquet sectors. Replication of the Floquet sectors produces an infinite ladder of energy replicas separated by $\Omega$, whose eigenvalues are identified as the quasi-energies of the Floquet Hamiltonian~\cite{hausinger2010dissipative}.

To further simplify the problem, we move to the interaction picture with respect to the modulated cavity frequency term \cite{hausinger2010dissipative, son2009floquet}, written as 
\begin{equation}
\label{eq:HsInRot}
H_S'(t)=g(e^{i\int_0^t \omega(t')dt'}a^\dagger\sigma + e^{-i\int_0^t \omega(t') dt'}a\sigma^\dagger )
\end{equation}
As periodic coefficients can be decomposed $e^{-i\int_0^t \omega(t')dt'}=\sum_n c_n e^{in\Omega t}$, the Floquet Hamiltonian in the interaction picture becomes
\begin{equation}
    \mathcal{H}_S' = g\, a^\dagger \sigma \otimes C^\dagger + g\, a \sigma^\dagger \otimes C + I \otimes \Omega N,
\end{equation}
where $C = \sum_{n} c_n\, T^n$ acts on the Floquet space. For the case $\omega(t) = -A \cos(\Omega t)$, the coefficients are given by $c_n = J_n(\frac{A}{\Omega})$, with $J_n(x)$ denoting the Bessel function of the first kind of order $n$.

The condition for coherent destruction of tunneling (CDT)~\cite{creffield2003location} in our scheme can be understood directly from the structure of the Floquet Hamiltonian. The term $I \otimes \Omega N$ in the Floquet Hamiltonian introduces an energy offset of $\Omega$ between adjacent Floquet sectors. As a result, in the high-frequency ($\Omega \gg g$) regime, transitions between different Floquet sectors are energetically suppressed, and the system dynamics is dominated by processes that occur within each Floquet sector. In this limit, the effective coupling between the TLS and the cavity mode is renormalized by the zero-order Bessel function, yielding an effective Hamiltonian of the form
\begin{equation}\label{eq:effHm}
    \mathcal{H}'_\text{eff,n} = (a^\dagger \sigma + a \sigma^\dagger) \otimes gJ_0\left(\frac{A}{\Omega}\right) |n\rangle\langle n|.
\end{equation}
Here, $g_{\text{eff}}=J_0\left(\frac{A}{\Omega}\right)g$ plays the role of an effective tunable coupling constant. Therefore, the CDT condition corresponds to $J_0\left(\frac{A}{\Omega}\right) = 0$, where the effective coupling vanishes, and the coherent energy exchange between the TLS and the cavity is suppressed. This mechanism enables dynamic control of the light-matter interaction by simply tuning the modulation parameters $A$ and $\Omega$.

\begin{figure}[h!]
    \centering
    \includegraphics[width=0.45\textwidth]{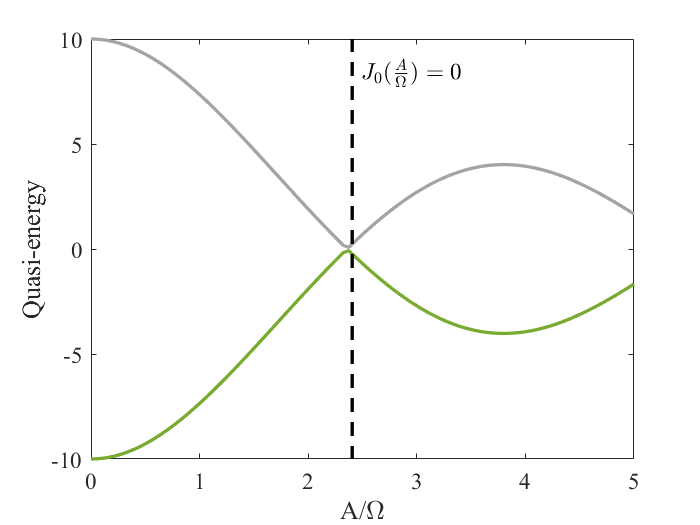}
    \caption{The dependence of two quasi-energies of the Floquet Hamiltonian on the modulation depth $A$. The modulation effectively renormalizes the cavity–TLS coupling strength according to the zero-order Bessel function $J_0(\frac{A}{\Omega})$. When $J_0(\frac{A}{\Omega})=0$, the Floquet eigenstates become degenerate, corresponding to the condition for CDT, where the TLS is effectively decoupled from the cavity mode.}
    \label{fig:bessel}
\end{figure}

Fig.~\ref{fig:bessel}(a) shows the dependence of the Floquet quasi-energy levels within a single Floquet sector on the ratio $\frac{A}{\Omega}$. In the high-frequency regime, the effective coupling constant between the TLS and the cavity is given by $J_0(\frac{A}{\Omega}) g$, so the energy splitting between the eigenstates is primarily determined by this renormalized coupling. When $J_0(\frac{A}{\Omega}) = 0$, corresponding to the CDT condition, the two levels approach near degeneracy. Physically, this implies that the period of Rabi oscillations diverges as the effective coupling vanishes. 

Building on the CDT analysis, the interaction on-off control demonstrated below is achieved by tuning the modulation amplitude $A$, which determines the effective coupling strength $g_{\text{eff}} = J_0\left(\frac{A}{\Omega}\right) g$. In particular, by varying $A$, the effective TLS-cavity coupling term $g_{\text{eff}}(a^\dagger \sigma + a \sigma^\dagger)$ can be continuously tuned from its bare value to zero, thereby enabling dynamical control of the coherent coupling. A key feature of our protocol is that the modulation amplitude $A$ is not simply switched abruptly between discrete values (e.g., 0 and 1). Instead, to guarantee phase preservation upon reactivation of the interaction, we implement the switching of $A$ using Gaussian-shaped rising and falling edges, which is given by

\begin{equation}\label{eq:Afloquet}
A(t) = 
A_0\Omega \begin{cases}
\exp\left(-\dfrac{(t - t_0)^2}{2\sigma_t^2}\right), &\ t < t_0 \\
1, &\ t_0 \leq t \leq t_1 \\
\exp\left(-\dfrac{(t - t_1)^2}{2\sigma_t^2}\right), &\ t > t_1
\end{cases}.
\end{equation}

Here, $t_0$ and $t_1$ denote the switching times and $\sigma_t$ controls the smoothness of the transition. This approach ensures a smooth transition between the interacting and noninteracting regimes, which is critical for maintaining high-fidelity quantum control and minimizing unwanted excitations \cite{deng2016dynamics}. During the interaction-off period, the system wavefunction can be approximated as 
\begin{equation}\label{eq:psiFloquet}
|\psi(t)\rangle = c_a(0) e^{i\int_0^t A(t')\cos(\Omega t')dt'} |1, g\rangle + c_e(0) |0, e\rangle, 
\end{equation}
where $A(t)$ captures the time-dependent modulation amplitude. Here, $|1, g\rangle$ denotes the state with one photon in the cavity and the TLS in its ground state, whereas $|0, e\rangle$ represents the vacuum-photon state with the TLS exciton excited. The coefficients $c_a(0)$ and $c_e(0)$ represent the initial amplitudes of the cavity mode and TLS exciton mode, respectively. In this regime, the relative phase of the cavity mode continues to accumulate according to the periodic drive, while the TLS component remains essentially unchanged. With a properly designed control sequence $A(t)$, the relative phase between the cavity and exciton components is maintained, allowing coherent Rabi oscillations to restart seamlessly when the coupling is restored.

\subsection{Environmental Decoherence: Bosonic Reservoir Coupling}

To account for environmental decoherence in realistic implementations, we extend the TLS-modulated cavity model by introducing a generic bosonic reservoir coupled to the two-level system. In solid-state realizations of light-matter interaction, such environmental coupling is typically dominated by lattice phonons \cite{roy2012polaron}. As a representative example, we therefore consider acoustic phonon coupling described by the spectral density in quantum dots system. The total Hamiltonian is then written as
\begin{equation}\label{eq:pntotH}
    H = H_S + H_E + H_I,
\end{equation}
where $H_S$ is the TLS-cavity Hamiltonian defined previously, $H_E = \sum_q \omega_q b_q^\dagger b_q$ describes the free phonon environment, and the interaction Hamiltonian stands for
\begin{equation}\label{eq:pnHI}
    H_I = \sigma^\dagger\sigma \sum_q \lambda_q \left( b_q^\dagger + b_q \right).
\end{equation}
Here $b_q^\dagger$ ($b_q$) is the creation (annihilation) operator for the $q$-th phonon mode and $\lambda_q$ is the coupling strength. To efficiently include multiphonon processes of the TLS-phonon coupling, we perform the polaron transformation on the total Hamiltonian. This transformation yields a renormalized system Hamiltonian together with a residual system-reservoir interaction term, whose explicit forms are provided in Appendix~\ref{app:polaron}.

To describe the non-Markovian dynamics induced by phonon interactions, we employ the Nakajima-Zwanzig (NZ) projection operator formalism \cite{roy2012polaron}. Within the Born approximation, i.e., to second order in the system–reservoir coupling, the NZ master equation for the reduced density operator $\tilde{\rho}(t)$ takes the form
\begin{multline}\label{eq:NZEqn}
\frac{\partial \tilde{\rho}(t)}{\partial t} = -i[\tilde{H}_S(t), \tilde{\rho}(t)] + \mathcal{L}(\tilde{\rho}) - \int_0^t R(t,t') dt',
\end{multline}
where $R(t,t')$ is the residual system–reservoir interaction term, as detailed in Appendix~\ref{app:polaron}. The dissipative dynamics of the cavity mode and the TLS exciton are modeled using the standard Lindblad operators,
\begin{align}
\mathcal{L}(\tilde{\rho})=&\frac{\kappa}{2}(2a\tilde{\rho} a^\dagger-a^\dagger a\tilde{\rho}-\tilde{\rho} a^\dagger a) \nonumber\\
+& \frac{\gamma}{2}(2\sigma\tilde{\rho} \sigma^\dagger-\sigma^\dagger \sigma\tilde{\rho}-\tilde{\rho} \sigma^\dagger \sigma),
\end{align}
where the $\kappa$ and $\gamma$ are the cavity and TLS decay rates, respectively.

\section{Results}

\subsection{Interaction Control}


\subsubsection{Bloch Sphere Representation}

\begin{figure}[h]
\centering
\includegraphics[width=0.4\textwidth]{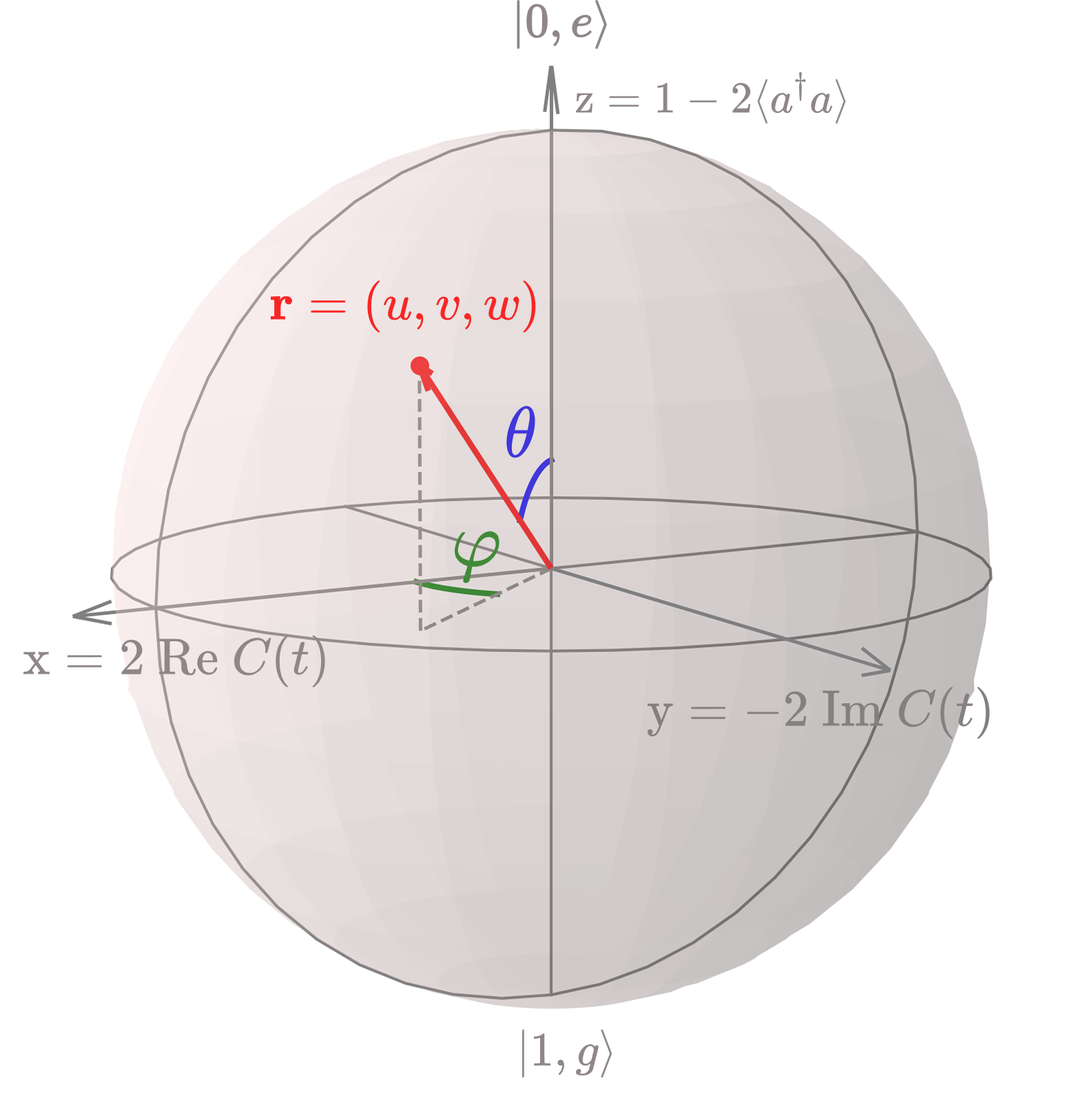}
\caption{
Bloch sphere representation of quantum states in the single-excitation subspace spanned by $\{|0,e\rangle,|1,g\rangle\}$. The Bloch vector $\mathbf r$ can be specified either by Cartesian coordinates $(u,v,w)$ or by spherical angles $(\theta,\varphi)$.
}
\label{fig:bloch-diagram}
\end{figure}

Within the single-excitation manifold, the state of the TLS-cavity system can be written as
\begin{equation}
|\psi(t)\rangle=c_e(t)|0,e\rangle+c_a(t)|1,g\rangle .
\end{equation}
As shown in Fig.~\ref{fig:bloch-diagram}, the Bloch vector $\mathbf r(t)=(u,v,w)$ is defined as
\begin{align}
u(t)&=2\,\mathrm{Re}\,C(t),\\
v(t)&=-2\,\mathrm{Im}\,C(t),\\
w(t)&=|c_e(t)|^2-|c_a(t)|^2,
\end{align}
where the coherence term is defined as $C(t)=\langle a^\dagger\sigma\rangle$.
In spherical coordinates, the state can be parametrized as
\begin{equation}\label{eq:bloch-state}
|\psi(t)\rangle
=
\cos\frac{\theta(t)}{2}\,|0,e\rangle
+
e^{i\varphi(t)}\sin\frac{\theta(t)}{2}\,|1,g\rangle,
\end{equation}
where $\theta(t)$ determines the population distribution between the two basis states, while $\varphi(t)$ characterizes their relative phase.


For resonant light-matter coupling, the Hamiltonian generates coherent Rabi oscillations corresponding to a rotation around the $x$ axis on the Bloch sphere. The trajectory therefore evolves in a fixed plane perpendicular to the $x$ axis, with its $x$-component $u$ conserved during the motion. For the common initial condition $u(0)=0$ (a state on the $y$-$z$ plane), the trajectory follows this meridian, where the polar angle $\theta(t)$ evolves in time and determines the population dynamics.

To characterize the orientation of the trajectory relative to the $x$ axis, we introduce an angle $\chi$ defined by 
\begin{equation}
\cos\chi(t)=\frac{u(t)}{|\mathbf r(t)|}.
\end{equation}
For resonant Rabi oscillations $u(t)$ remains constant (for a pure state
$|\mathbf r|=1$), and therefore $\chi$ is conserved during the evolution.

The quantities $\theta$ and $\chi$ will therefore be used to characterize
phase preservation under different interaction-control protocols.

\subsubsection{Phase-Preserving Mechanism}

To preserve the oscillation phase, the control protocol should effectively freeze the net phase accumulation associated with the switching operation, so that no phase offset remains once the modulation is turned off.  In essence, the modulation envelope must vary smoothly compared to the modulation frequency $\Omega$ to maintain phase coherence. This is achieved by introducing Gaussian-shaped rising and falling edges, as discussed previously. The most relevant parameters are the modulation frequency $\Omega$ and the durations of the rising and falling edges $\sigma_t$. 

We first clarify the phase behavior in a rotating basis adapted to the time-dependent cavity frequency. 
We introduce the basis
\begin{equation}
\{|0,e\rangle,\ e^{i\Phi(t)}|1,g\rangle\},
\end{equation}
where
\begin{equation}\label{eq:PhiDef}
\Phi(t) = -\int_0^t \omega(t')\,dt',
\qquad
\omega(t) = -A(t)\cos(\Omega t),
\end{equation}
as defined in the main text. 
In this representation, the fast frequency modulation is absorbed into the time-dependent phase factor $e^{i\Phi(t)}$.

Under the condition that the envelope $A(t)$ varies slowly compared with the modulation frequency $\Omega$, the total phase accumulated in this rotating-frame transformation is
\begin{equation}
\Phi(+\infty)
=
\int_{0}^{+\infty} A(t)\cos(\Omega t)\,dt.
\end{equation}
For the smoothly varying envelopes considered in our protocol and in the
high-frequency $\Omega$ regime, as shown in Appendix~\ref{app:rot_phase}, we obtain
\begin{equation}
\Phi(+\infty) \approx 0 .
\end{equation}

This result has a clear interpretation. The time-dependent basis transformation $|1,g\rangle \rightarrow e^{i\Phi(t)}|1,g\rangle \rightarrow |1,g\rangle$ forms a closed cycle over the entire switching process and returns to the original reference frame at long times. Consequently, the rotating-frame transformation itself does not introduce any residual phase between $|0,e\rangle$ and $|1,g\rangle$ once the interaction is restored.

\begin{figure}[h]
    \centering
    \includegraphics[width=0.5\textwidth]{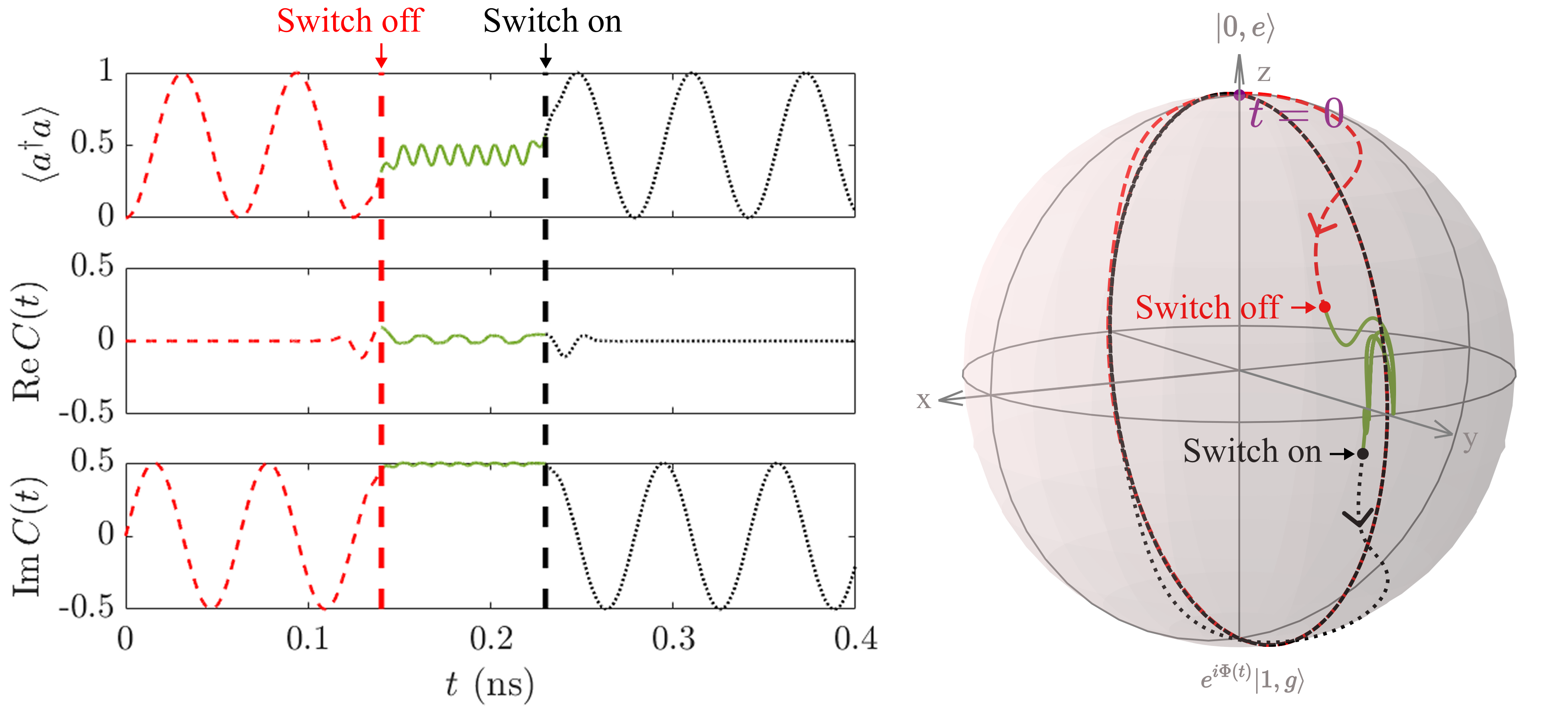}
    \caption{{\color{red}
    The Floquet control scheme. 
Left: Population $\langle a^\dagger a \rangle$ and coherence 
$C(t)=e^{i\Phi(t)}\langle a^\dagger \sigma \rangle$ (real and imaginary parts), 
whose argument determines the relative phase of wavefunction $|\psi(t)\rangle$ between 
$|0,e\rangle$ and $e^{i\Phi(t)}|1,g\rangle$. 
Right: Corresponding Bloch sphere trajectory.    
    The calculations are performed with parameters $A_0\approx 2.4\Omega$, $\Omega/(2\pi)=50\,\mathrm{GHz}$, $g/(2\pi)=8\,\mathrm{GHz}$, $\sigma_t=0.02 \mathrm{ns}$, where $A_0$ is chosen to minimize the effective coupling. Unless otherwise specified, these parameters are used throughout the following calculations.}
    }
    \label{fig:floquet-bloch}
\end{figure}

The above rotating-frame argument explains why no phase is introduced by the basis transformation itself after total switching process. To further understand the dynamical mechanism dominated by $H_S'(t)$ in Eq.~\ref{eq:HsInRot}, we examine the evolution on the Bloch sphere within the single-excitation subspace. The dynamics is illustrated in Fig.~\ref{fig:floquet-bloch}, where the left panel shows the population $\langle a^\dagger a\rangle$ and the coherence $C(t)$, while the right panel displays the corresponding Bloch sphere trajectory. For the envelope $A(t)$ in Eq.~\eqref{eq:Afloquet}, the effective switch-off and switch-on times in the figures are taken as $t_0-\sigma_t$ and $t_1+\sigma_t$ to account for the smooth Gaussian switching edges. The coherence is defined as $C(t)=e^{i\Phi(t)}\langle a^\dagger\sigma\rangle$ in the rotating basis, which reflects the dynamics governed by the effective Hamiltonian as Eq.~\ref{eq:HsInRot}.

In the absence of modulation, the Hamiltonian generates resonant Rabi oscillations around the $x$-axis, corresponding to motion of the Bloch vector along a meridian in the $y$-$z$ plane. The initial state $|0,e\rangle$ is located at the north pole, and the dynamics proceeds as a uniform rotation around the fixed $x$ direction. In this regime, the population and the imaginary part of the coherence oscillate sinusoidally, while $\mathrm{Re}\,C(t)$, which is the $x$-component remains zero, reflecting the circular motion of the Bloch vector in the $y$-$z$ plane.

The switching protocol is implemented through a periodically modulated frequency term $\omega(t)=-A(t)\cos(\Omega t)$ with a slowly varying envelope $A(t)$ defined in Eq.~(\ref{eq:Afloquet}). The envelope consists of three regions: a Gaussian rising edge for $t<t_0$, a plateau with constant amplitude for $t_0\le t\le t_1$, and a Gaussian falling edge for $t>t_1$. Far away from the switching window, the envelope approaches zero ($A(t)\rightarrow0$), and the Hamiltonian reduces to the resonant light-matter interaction, leading to ordinary Rabi oscillations. As shown in Fig.~\ref{fig:floquet-bloch}, long before the switching point $t_0$ the Bloch vector (red dashed trajectory) undergoes a uniform rotation in the $y$-$z$ plane. As the system enters the rising edge near $t_0$, the modulation amplitude gradually increases and the instantaneous rotation axis starts to deviate from the static $x$ direction on the Bloch sphere. Consequently, the Bloch trajectory begins to exhibit small oscillatory deviations from the original meridian. During the plateau region ($t_0<t<t_1$), the modulation reaches its maximum strength and induces a rapidly varying effective coupling as in Eq.~\ref{eq:HsInRot}. The Bloch trajectory therefore develops a bounded micromotion and forms a small Lissajous-like pattern around a fixed point, reflecting the higher-order Floquet micromotion. Finally, in the falling edge near $t_1$, the envelope $A(t)$ decreases smoothly back to zero, restoring the resonant interaction. On the Bloch sphere, the trajectory gradually leaves the micromotion orbit and smoothly oscillates back to the original meridian. This behavior reflects the fact that the modulation does not introduce a persistent component of the effective rotation axis away from the $x$ direction. As a result, although the Bloch vector may undergo small excursions
during the switching duration, its $x$ projection returns to the same value once the envelope $A(t)$ vanishes. Consequently, when the modulation is turned off, the system resumes the same Rabi oscillation with the phase reference $\chi$ preserved.

\begin{figure}[h]
  \centering
  \includegraphics[width=0.45\textwidth]{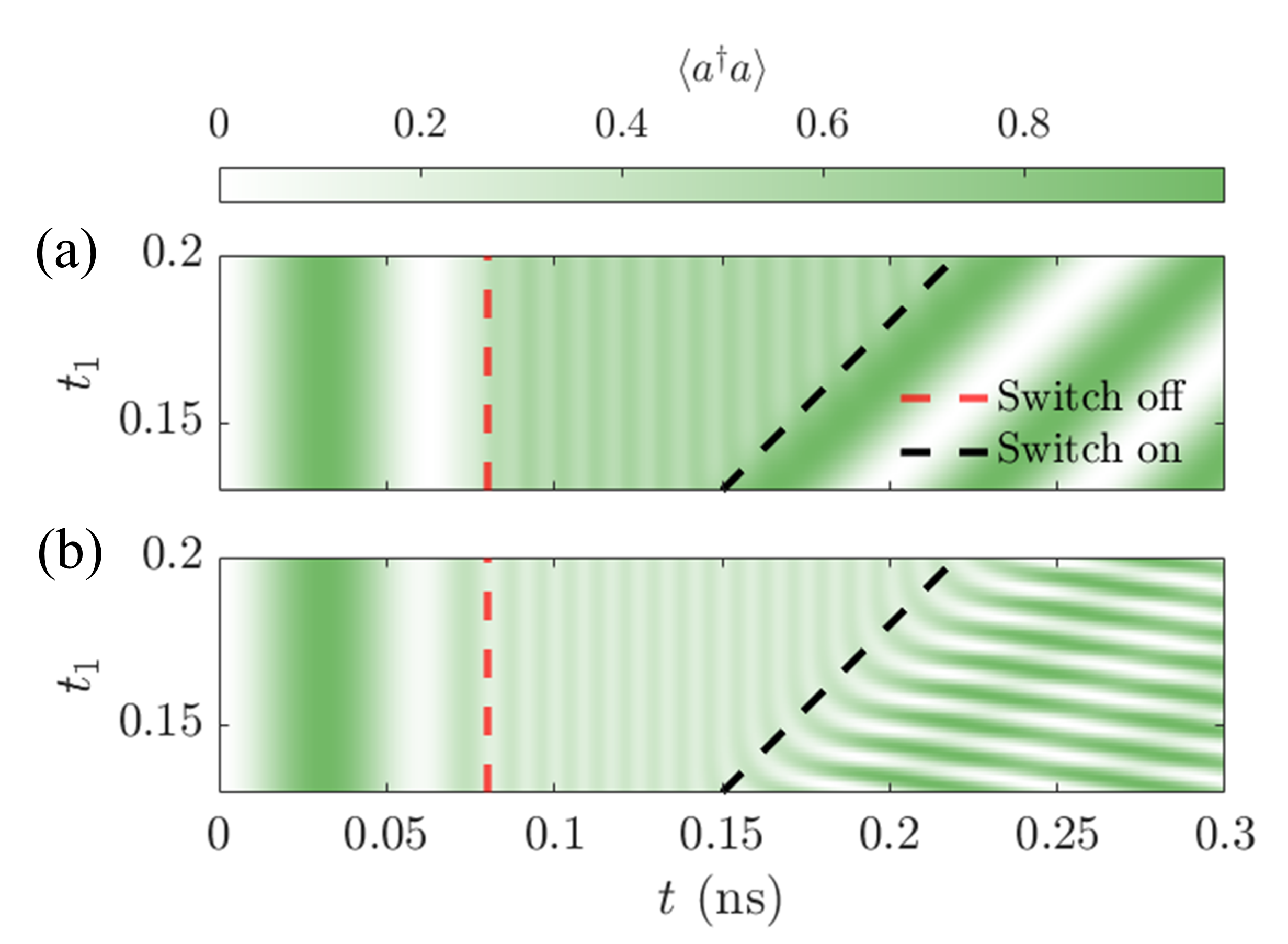}
  \caption{Comparison of cavity population dynamics under different interaction-control protocols. The cavity population is shown when the interaction is switched off (red dashed) and later reactivated (black dashed) at various times. For static detuning-based control (b), the reactivation dynamics depends periodically on the interruption time due to continuous phase accumulation during the detuned period. In contrast, for Floquet-based control (a), the trajectories remain identical regardless of interruption time, confirming robust phase preservation.}
  \label{fig:TendScan}
\end{figure}

In addition, Fig.~\ref{fig:TendScan}(a) shows that the population dynamics after reactivation is insensitive to the switching duration $t_1-t_0$. This indicates that the polar angle $\theta$ does not acquire an additional offset during the plateau. The mismatch between the switch-off and switch-on points in Fig.~\ref{fig:floquet-bloch} is determined by the rising and falling edges, where a finite effective coupling $g_{\mathrm{eff}}$ remains present. This intuitive picture is consistent with the analytical result obtained from a high-frequency Magnus expansion of the evolution operator in Appendix~\ref{app:dyn_magnus}. Under the slowly varying envelope condition, the rapidly oscillating modulation-induced terms average out over the cycles, and the leading effective Hamiltonian retains only the resonant component proportional to $a^\dagger\sigma + a\sigma^\dagger$, up to corrections of order $O(1/\Omega)$. Consequently, the dynamics before and after the switching reduces to a Rabi rotation around the $x$ axis.

This behavior can also be understood within the adiabatic Floquet framework \cite{berry1984quantal,deng2016dynamics,ying2020geometric}. For a slowly varying envelope $A(t)$, the system follows instantaneous Floquet modes with quasi-energies $\varepsilon_\pm(A)$. 
The total phase accumulated during the switching process consists of dynamical and geometric contributions. For the present driving protocol the geometric phase vanishes \cite{deng2016dynamics,ying2020geometric}, while under the CDT condition the quasi-energies $\varepsilon_\pm(A)$ approach zero in the plateau region ($t_0<t<t_1$), so that the dynamical phase accumulated there is negligible. Consequently, the switching process does not introduce an additional phase shift, and the change in $\theta$ originates from the rising and falling edges of the modulation.

\begin{figure}[h]
    \centering
    \includegraphics[width=0.5\textwidth]{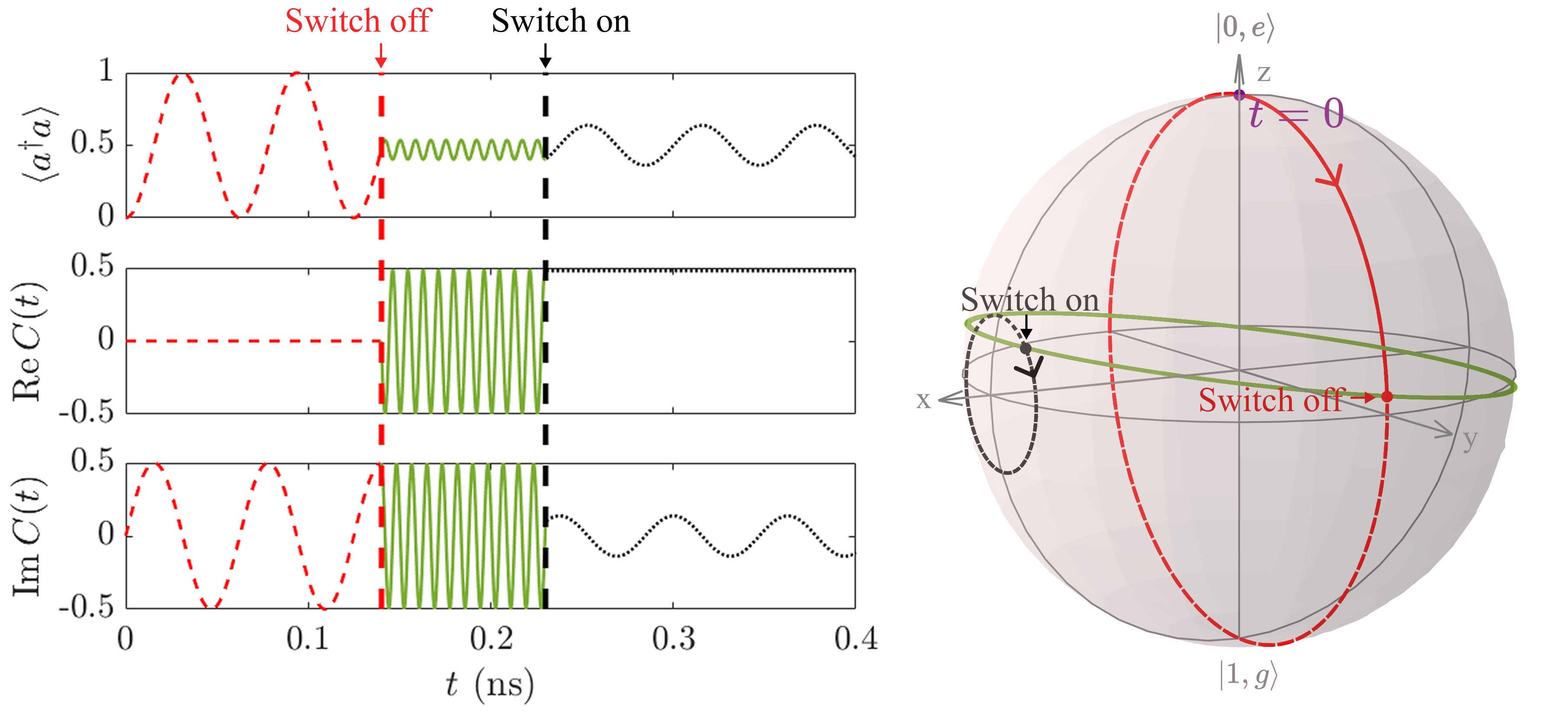}
    \caption{
The static detuning control scheme. 
Left: Population $\langle a^\dagger a \rangle$ and coherence 
$C(t)=\langle a^\dagger \sigma \rangle$ (real and imaginary parts), 
whose argument determines the relative phase between 
$|0,e\rangle$ and $|1,g\rangle$. 
Right: Corresponding Bloch sphere trajectory.
The phase of the oscillations is not preserved when the interaction switched off at time $t_0$ and reactivated at time $t_1$ in the static cavity detuning scheme. The maximum modulation depth $A_0$ takes same maximum modulation depth with Fig.~\ref{fig:floquet-bloch}.}
\label{fig:static-bloch}
\end{figure}

For comparison, we also examine a widely used interaction-control approach based on static cavity detuning \cite{ridolfo2011all, faraon2010fast}, where a large detuning is applied as $\omega(t)=A_0\Omega$ for $t \in [t_0, t_1]$ and vanishes otherwise. In this scheme, the Hamiltonian acquires a finite detuning term proportional to $a^\dagger a$, which tilts the instantaneous rotation axis from the $x$ direction into the $x$-$z$ plane during the switching duration (see Fig.~\ref{fig:static-bloch}). As a result, between $t_0$ and $t_1$, the state no longer evolves along the original meridian associated with resonant Rabi oscillation, but instead precesses around the tilted axis. Consequently, the Bloch vector departs from the initial meridian and both the polar and azimuthal coordinates evolve continuously during the detuned interval, as illustrated by the blue trajectory. When the detuning is switched off and the rotation axis returns to the $x$ direction, the Bloch vector generally does not return to the original meridian, but instead acquires a shift in both $\chi$ and $\theta$. The subsequent evolution therefore follows a shifted orbit, shown as the dotted black trajectory after $t_1$, leading to a phase mismatch in the two-level coherence relative to the original trajectory. As a result, even with smooth switching $\omega(t)=A(t)$ in Eq.~\ref{eq:Afloquet}, the static detuning scheme generally produces a phase mismatch dependent on the switching duration as shown in Fig.~\ref{fig:TendScan}.

In both the static detuning and Floquet control schemes, small-amplitude oscillations can be observed in the cavity population during the interaction-off period. Although these oscillations do not affect the preservation of the phase at the control points, their physical origins are distinct in the two schemes. In the static detuning scheme, the state after switching-off is written as $|\psi(t)\rangle = c_{+} e^{-i\omega_{+} t} |\psi_{+}\rangle + c_{-} e^{-i\omega_{-} t} |\psi_{-}\rangle$, resulting in the cavity population $|\langle g,1 | \psi(t) \rangle|^2 = |c_{+}\langle g,1|\psi_{+}\rangle e^{-i\omega_{+} t} + c_{-}\langle g,1|\psi_{-}\rangle e^{-i\omega_{-} t}|^2$. The large eigen frequency difference $\tilde{\Delta} = \omega_{+}-\omega_{-}$ leads to an interference term, producing oscillations with relative phases accumulated linearly in time. As a result, the phase at the next control point depends sensitively on the exact duration and cannot, in general, be preserved except at special revival times $t_1-t_0 = 2\pi m / \tilde{\Delta}$. In the Floquet scheme, the eigenstates $|u_{\pm}(t)\rangle$ are time-periodic with quasi-energies $\varepsilon_{\pm}$. Near the CDT point $J_0(A/\Omega) = 0$, the quasi-energy splitting is strongly suppressed such that $\varepsilon_{+} \approx \varepsilon_{-}$. The residual oscillations in the cavity population therefore originate not from large energy differences but from the intrinsic time-periodic coefficients of the Floquet modes themselves. 

\subsubsection{Switching Rate Limit}
\begin{figure}[h]
    \centering
    \includegraphics[width=0.45\textwidth]{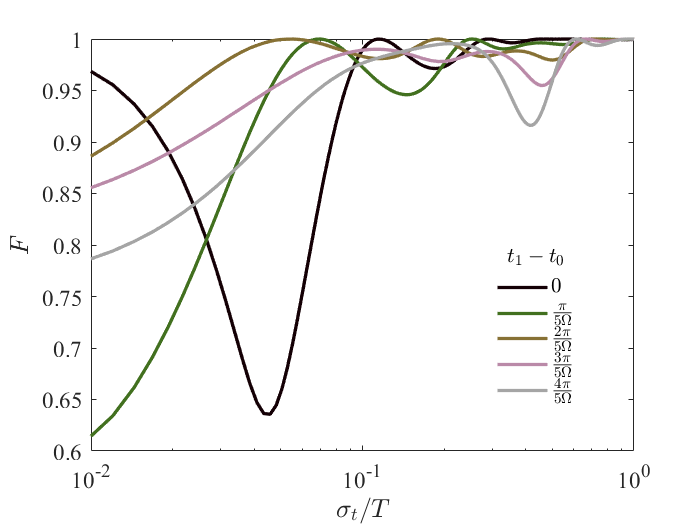}
    \caption{Fidelity $F$ as a function of the normalized switching time $\sigma_t/T$. Here, $t_0 = 20\pi/\Omega$ is fixed while the duration of the control operation is varied.
    }
    \label{fig:FidelitySigmaOmega}
\end{figure}

We define the fidelity of the control protocol as $F=\max_{t>t_1} |\langle g,1| \psi(t)\rangle|$. In the ideal case $F$ reaches unity, whereas the control operation perfectly freezes and restores the system dynamics without inducing unwanted transitions or phases, reflecting the complete recoverability of the coherent interaction. The fidelity $F$ exceeds 0.99 when $\sigma_t \gtrsim 0.6T$, for an arbitrary switching-off duration $t_1-t_0$, indicating robust control in the adiabatic regime, as shown in Fig.~\ref{fig:FidelitySigmaOmega}.

\begin{figure}[h]
    \centering
    \includegraphics[width=0.45\textwidth]{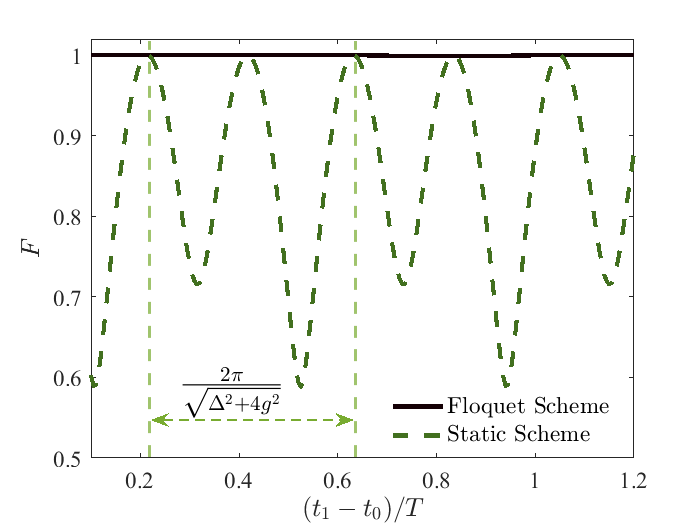}
    \caption{Fidelity $F$ as a function of the normalized duration time $(t_1-t_0)/T$ for Floquet scheme and static detuning scheme. Here, the interruption time is fixed at $t_0 = 20\pi/\Omega$, while the restart time $t_1$ of the control operation is varied.
    }
    \label{fig:FidelityDuration}
\end{figure}

Fig.~\ref{fig:FidelityDuration} compares the fidelity $F$ as a function of the normalized interaction-off durations $(t_1-t_0)/T$ for the Floquet and static detuning schemes. For the Floquet scheme, $F$ remains nearly constant over a wide range of durations, indicating insensitivity to interaction-off durations. In contrast, the static detuning scheme exhibits oscillations in $F$ with a period $2\pi/\tilde{\Delta}$, where $\tilde{\Delta} = \sqrt{\Delta^2 + 4g^2}$ is the generalized Rabi frequency for the static detuning system. This reflects a periodic rotation of the state on the Bloch sphere, leading to revival of fidelity at integer multiples of the oscillation period. Compared with the Floquet scheme, the static detuning scheme requires precise control over the duration of the interaction-off period to achieve high fidelity, thereby reducing the robustness of the control protocol.

\begin{figure}[h]
    \centering
    \includegraphics[width=0.45\textwidth]{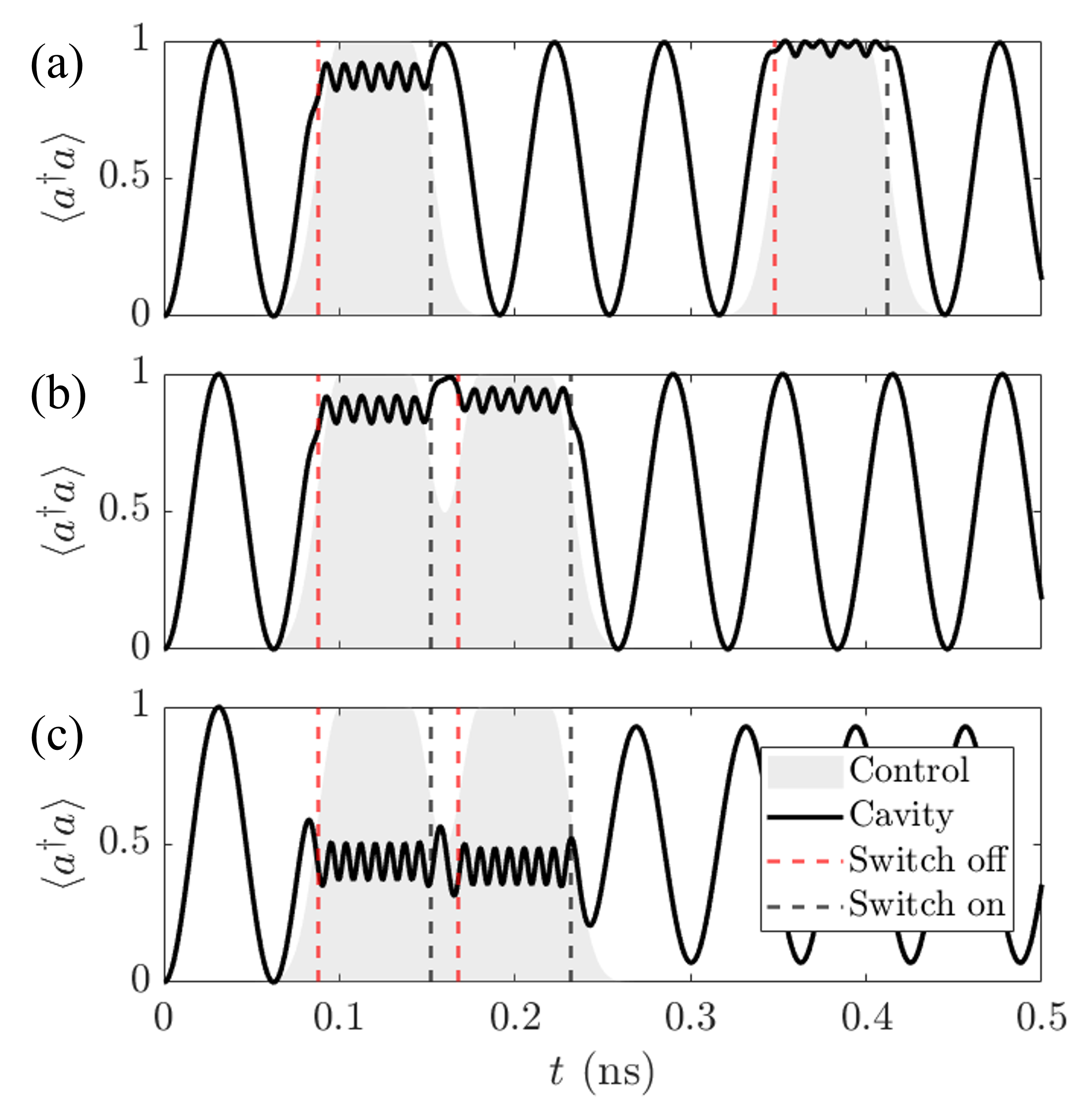}
    \caption{
    Dynamics of the cavity population under two successive off-on control sequences in the Floquet control scheme, where the gray shades represent the modulation amplitude $A(t)$.
    (a) The two control windows are well separated in time, yielding identical phase preservation as in the single-sequence case. 
    (b) The two control windows partially overlap, leading to a short segment of Rabi oscillations during the intermediate switch-on to switch-off transition. 
    In both cases, the phase at the control points is preserved.
    (c) Static-detuning counterpart under the same successive switching protocol, where the population oscillation shows a phase shift relative to the original trajectory and a reduced maximum population.}
    \label{fig:PopulationRepeat}
\end{figure}

Furthermore, the robustness of the control protocol must also be evaluated, particularly in cases involving successive switch off-on operations. Fig.~\ref{fig:PopulationRepeat} illustrates the effect of applying successive switch off-on control sequences. When multiple off-on control windows with a long or short interval are applied in sequence, the phase is preserved in either case, demonstrating the robustness of the proposed scheme. According to Eq.~\eqref{eq:phase_f}, as long as the control signal varies sufficiently slowly, the accumulated phase during the switching process remains negligible. This explains why phase preservation is maintained even when the two control windows partially overlap.

\subsection{Phonon-induced Dephasing}
In semiconductor quantum emitters, coupling to lattice phonons is the dominant source of pure dephasing and gives rise to non-Markovian coherence loss~\cite{esmann2024solid, kaer2010non, morreau2019phonon}. Since the Floquet control protocol relies on maintaining a well-defined exciton-photon phase relation during the interaction-off interval, it is essential to determine whether phonon-induced decoherence compromises this phase-preserving feature. To quantitatively investigate the influence of phonon-induced dephasing on the interaction control protocol~\cite{roy2012polaron}, we numerically solve the NZ master equation Eq.~(\ref{eq:NZEqn}). In this approach, the TLS-cavity subsystem can be conveniently represented using a truncated basis consisting of the lowest three states, $|1\rangle \equiv |0,e\rangle$, $|2\rangle \equiv |1,g\rangle$, and $|3\rangle \equiv |g,0\rangle$~\cite{kaer2010non,kaer2012microscopic,morreau2019phonon}. This choice is justified in the regime of low photon number, where higher excited states are negligibly populated.

\begin{figure}[t]
    \centering
    \includegraphics[width=0.95\linewidth]{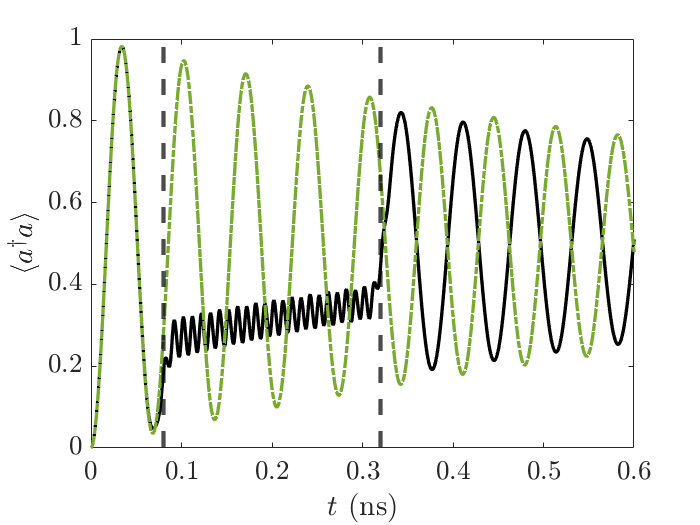}
    \caption{Time evolution of the TLS-cavity system under phonon-induced dephasing without considering other decay and dephasing channels. Black solid lines correspond to the interaction-control case (with switching off and on), while green dashed lines show the unmodulated reference case. The phonon parameters taken into calculations are $\alpha_p = 0.06~\text{ps}^2$, $T=4~\text{K}$ and $\omega_b = 1~\text{meV}$ \cite{roy2012polaron,mccutcheon2010quantum}.}
    \label{fig:floquetcoherencewpn}
\end{figure}

To assess the role of phonons during Floquet-based interaction control, we compare two distinct scenarios, one case in which the TLS-cavity interaction is temporarily switched off and subsequently restored, and a reference case with continuous, unmodulated interaction. Phonon-induced effects in these two scenarios are illustrated in Fig.~\ref{fig:floquetcoherencewpn}. From numerical results, it is evident that dephasing persists during the interaction-off period and the population maxima after reactivation exhibit only minor changes, demonstrating that switching off the interaction does not substantially modify the overall phonon-induced decay.

For a more quantitative characterization of the decoherence process, we derive an explicit expression for the phonon-induced dephasing rate within the polaron NZ framework. In the non-Markovian NZ master equation Eq.~(\ref{eq:NZEqn}), the relevant contribution arises from the second-order kernel $R(t,t')$.

In the single-excitation subspace, we expand the system density matrix $\rho(t')$ as $\rho_{ij}(t')$, and denote the evolution operator as $U_S(t,t')$ with elements $u_{ij}$. The evolution of the TLS-cavity coherence is governed by the off-diagonal component $R_{21}$, whose coefficient acting on $\rho_{21}$ yields the effective dephasing rate. The corresponding term can be written as
\begin{equation}\label{eq:deph_pn}
R_{21}^{(\mathrm{deph})} = -2g^2\, \left(2|u_{21}|^2 - 1 \right)\mathrm{Re} \left[\langle B \rangle^2 \left(e^{\phi} - 1 \right)\right],
\end{equation}
where $\langle B \rangle$ is the thermal expectation value of the phonon displacement operators, and $\phi$ is the phonon correlation function. The rotating wave approximation (RWA) is used to obtain the expression of $|u_{21}|^2$, which can be written as $u_{21}(t,t')=-i\sin(gJ_0(\frac{A}{\Omega})(t-t'))\exp(-i\frac{A}{\Omega}\sin(\Omega t)+i\frac{A}{\Omega}\sin(\Omega t'))$ \cite{hausinger2010dissipative}. As $|u_{21}|^2\approx 0$ for the short time interval determined by the phonon correlation time, the contribution of the modulation-dependent term becomes negligible. In this regime, the dephasing rate is therefore dominated by the phonon correlation function and the coupling strength. This analysis confirms that, for $g \ll \Omega$, the Floquet modulation of the light–matter coupling has little impact on the intrinsic phonon-induced dephasing, which remains primarily governed by the TLS–phonon interaction itself~\cite{nazir2008photon}.

\begin{figure}[h!]
    \centering
    \includegraphics[width=0.45\textwidth]{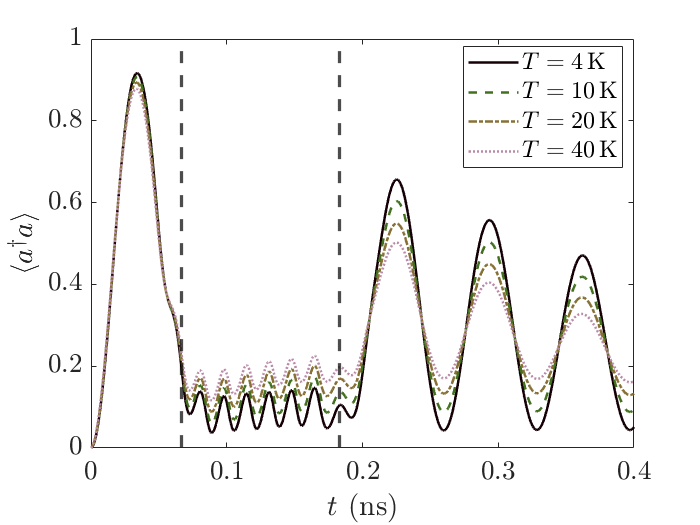}
    \caption{Time evolution of the cavity mode population at different temperatures. Increasing temperature leads to stronger phonon-induced dephasing, which suppresses the oscillations during the switching-off interval while leaving the phase and timing of oscillations largely unaffected.
    }
    \label{fig:phonon}
\end{figure}

The temperature dependence of the dynamics is summarized in Fig.~\ref{fig:phonon}, where we include cavity and TLS decay with rates $\kappa=4~\text{GHz}$ and $\gamma=1~\text{GHz}$, respectively. For a direct comparison with identical oscillation periods, we set $\langle B \rangle = 1$ in the system Hamiltonian of Eq.~(\ref{eq:tildeHS}), so that temperature only affects the dynamics through the system–reservoir interaction term. As temperature increases, phonon-induced dephasing suppresses the oscillation amplitude and damps the high-frequency micromotion associated with higher-order Floquet modes. Importantly, the positions of the oscillation extrema remain unchanged following the reactivation of TLS-cavity coupling, demonstrating that the phase and timing of the control sequence are robust against thermal fluctuations.

\section{Conclusion}

In this work, we have developed a Floquet-engineered scheme for coherent control of light-matter interactions in a TLS-modulated cavity system. By introducing smoothly varied modulation envelopes, the scheme demonstrates a phase-preserving feature during the interruption and reactivation of the control process. These insights pave the way for practical applications in photonic quantum gates, coherent quantum memories, and hybrid quantum devices, offering new strategies for scalable and robust quantum information processing on solid-state platforms.

\acknowledgements{This work is supported by the Zhejiang Province Leading Geese Plan (2024C01105), the National Future Industry Innovation Mission of China, the Beijing Natural Science Foundation (L248103), the National Key Research and Development Program of China (2021YEB2800500), and the National Natural Science Foundation of China (61574138, 61974131).}

\appendix

\section{Effective Rotating Phase}
\label{app:rot_phase}

The geometric phase accumulated during the interaction-off process \cite{deng2016dynamics, ying2020geometric}, corresponding to the bare-cavity phase evolution in Eq.~\eqref{eq:psiFloquet}, is obtained from Eq.~\eqref{eq:PhiDef} by taking the limit $t\to+\infty$,
\begin{equation}
    \Phi = \int_{0}^{+\infty} A(t)\cos(\Omega t)\,dt .
\end{equation}
To analyze the phase dynamics in detail, we decompose the integral into three contributions corresponding to the rising edge ($\Phi_r$), the plateau ($\Phi_m$), and the falling edge ($\Phi_f$) of the modulation envelope. For the plateau region, where $A(t) = A_0 \Omega$, the accumulated phase is given by $\Phi_m=A_0 \left( \sin(\Omega t_1) - \sin(\Omega t_0) \right)$.
For the falling edge, the analytic expression reads $
    \Phi_f=A_0\Omega\sqrt{\frac{\pi }{2}} \sigma_t  e^{-\frac{1}{2} \sigma_t ^2 \Omega ^2} \left(\cos (\Omega t_1)-\text{erfi}\left(\frac{\sigma_t  \Omega }{\sqrt{2}}\right) \sin (\Omega t_1)\right)\approx -A_0\sin(\Omega t_1)$.
The approximation holds when $\sigma_t\Omega \gg \sqrt{2}$. The rising edge is treated analogously, yielding $\Phi_r\approx A_0\sin(\Omega t_0)$. Summing these contributions, in the $\sigma_t\Omega \gg \sqrt{2}$ condition, we find analytically
\begin{equation}\label{eq:0phase}
I = \Phi_r + \Phi_m + \Phi_f \approx 0,
\end{equation}
demonstrating that the total phase accumulated during the switching process vanishes. This result implies that, as long as the edge durations are at least several modulation periods, the additional phase accrued during these transitions is negligible and the relative phase between the cavity and TLS states is effectively preserved. 

More generally, the choice of the rising and falling edges is not restricted by a Gaussian-shaped waveform. Eq.~(\ref{eq:0phase}) holds for any sufficiently slow switching protocol with variation rates much lower than $\Omega$, indicating that the phase accumulation in the bare state basis vanishes whenever the transition regions are adiabatic on the timescale of modulation. To see this more explicitly, a general smooth switching function for the falling edge is considered as $A(t) = A_0\Omega f_1(t)$ for $t > t_1$, with $f_1(t_1)=1$. The corresponding integral of the phase variation $\Phi_f=\int_{t_1}^{\infty} A_0\Omega f_1(t)\cos(\Omega t)dt$, yields
\begin{equation}\label{eq:phase_f}
\Phi_f = -A_0 f_1(t_1) \sin(\Omega t_1) - \frac{A_0}{\Omega} f_1'(t_1) \cos(\Omega t_1) + O\left(\frac{1}{\Omega^2}\right).
\end{equation}
If smoothness condition $A_0 f_1^{(n)}(t) \ll \Omega^n$ is satisfied, as is typically the case for sufficiently smooth decaying functions such as exponential, we have $\Phi_f \approx -A_0\sin(\Omega t_1)$. An analogous result holds for the rising edge. Thus, for any smooth and sufficiently slow switching protocol, the total accumulated geometric phase is strongly suppressed, with the residual contribution scaling as $O\left(\frac{1}{\Omega}\right)$. This demonstrates the robustness of the phase-preserving property against variations in the specific choice of switching envelope.

\section{Effective Dynamics from the Magnus Expansion}
\label{app:dyn_magnus}

To understand the phase-preserving switching dynamics, we analyze the interaction Hamiltonian
\begin{equation}
H_S'(t)=g\left(e^{-i\Phi(t)}a^\dagger\sigma + e^{i\Phi(t)}a\sigma^\dagger\right),
\end{equation}
with
\begin{equation}
\Phi(t)=\int_0^t A(t')\cos(\Omega t')\,dt'.
\end{equation}

For a smoothly varying envelope $A(t)$, integration by parts gives
\begin{equation}
\Phi(t)\approx \frac{A(t)}{\Omega}\sin(\Omega t),
\end{equation}
up to corrections of order $O(\dot A/\Omega^2)$. Defining $\beta(t)=A(t)/\Omega$, the Jacobi-Anger identity yields
\begin{equation}
e^{\pm i\beta(t)\sin(\Omega t)}
=
\sum_{n=-\infty}^{\infty} J_n(\beta(t))e^{\pm in\Omega t}.
\end{equation}

Substituting this expansion into the Hamiltonian leads to
\begin{equation}
H_S'(t)
=
\sum_{n} gJ_n(\beta(t))
\left(
a^\dagger\sigma e^{-in\Omega t}
+
a\sigma^\dagger e^{in\Omega t}
\right).
\end{equation}

The $n=0$ component is
\begin{equation}
H_0(t)=gJ_0(\beta(t))(a^\dagger\sigma+a\sigma^\dagger).
\end{equation}

Using the Magnus expansion $U(t)=\exp(\Omega_1+\cdots)$, the first-order term reads
\begin{equation}
\Omega_1=-i\int H(t)\,dt.
\end{equation}
Because the terms with $n\neq0$ contain rapidly oscillating factors $e^{in\Omega t}$, their contributions are suppressed by the fast modulation and scale as $O(g/\Omega)$. The leading contribution therefore arises from the $n=0$ component,
\begin{equation}
\Omega_1
\approx
-i\int gJ_0(\beta(t))(a^\dagger\sigma+a\sigma^\dagger)\,dt .
\end{equation}

The accumulated Rabi rotation angle during a switching interval $[t_0,t_1]$ is therefore
\begin{equation}
\Theta_x=\int_{t_0}^{t_1} gJ_0(\beta(t))\,dt .
\end{equation}
In the CDT regime, where $J_0(\beta)\approx0$, this rotation angle vanishes to leading order, indicating that no Rabi oscillation is accumulated during the interaction-off window.

Higher-order Magnus terms correspond to virtual processes between different modulation harmonics and are suppressed by the fast modulation, scaling as $O(g^2/\Omega)$ with additional reduction from the slow envelope variation. Consequently, the effective interaction retains the same operator structure as the resonant Jaynes-Cummings coupling and does not generate a persistent rotation-axis tilt.

\section{Polaron Master Equation}
\label{app:polaron}

We apply the polaron transformation to the total Hamiltonian in Eq.~(\ref{eq:pntotH}) \cite{kaer2012microscopic,roy2012polaron}, where the TLS-phonon interaction is given by Eq.~(\ref{eq:pnHI}). The transformation is defined by the unitary operator
\begin{equation}
    U_{\text{pol}} = \exp \left[ \sigma^\dagger \sigma \sum_q \frac{\lambda_q}{\omega_q} (b_q^\dagger - b_q) \right].
\end{equation}
Applying this transformation yields
\begin{equation}
    \tilde{H}(t) = U_{\text{pol}} H U^{-1}_{\text{pol}} = \tilde{H}_S(t) + \tilde{H}_I + \tilde{H}_E,
\end{equation}
with the three components representing, respectively, the transformed system Hamiltonian, the residual phonon-induced interaction, and the free phonon modes $\tilde H_E = \sum_q \omega_q b_q^\dagger b_q$.

The transformed system Hamiltonian takes the form
\begin{equation}\label{eq:tildeHS}
    \tilde{H}_S(t) = \omega(t)\, a^\dagger a + g \langle B \rangle (a^\dagger \sigma + \sigma^\dagger a) + \Delta_{\mathrm{pol}}\, \sigma^\dagger\sigma,
\end{equation}
where $g \langle B \rangle$ is the renormalized light–matter coupling and $\Delta_{\mathrm{pol}}$ is the polaron shift. Since the TLS resonance can be redefined to incorporate $\Delta_{\mathrm{pol}}$, we set $\Delta_{\mathrm{pol}}=0$ in simulations. The residual interaction \cite{roy2012polaron} with the phonon reservior is captured by
\begin{align}
    \tilde{H}_I &= X_g\xi_g + X_e\xi_e, \\
    X_g &= g(a^\dagger \sigma + \sigma^+ a), \label{eq:Xg}\\
    X_e &= ig(\sigma^+ a - a^\dagger \sigma), \label{eq:Xe}\\
    \xi_g &=\frac{1}{2}(B_++B_--2\langle B \rangle), \label{eq:Xig}\\
    \xi_e &=\frac{1}{2i}(B_+-B_-), \label{eq:Xie}
\end{align}
where the coherent displacement operators of the phonon modes are 
\begin{align}
B_{\pm}=\exp\left[\pm\sum_q\frac{\lambda_q}{\omega_q}(b_q-b^\dagger_q) \right].
\end{align} 

In thermal equilibrium, the phonon fluctuations satisfy
\begin{equation}
    \langle \xi_m(t)\,\xi_n(t') \rangle_B
    = \delta_{mn}\, G_m(t-t'), \quad m,n \in \{g,e\},
\end{equation}
with
\begin{align}
    G_g(\tau) &= \langle B\rangle^2 \left[ \cosh\phi(\tau) - 1 \right], \\
    G_e(\tau) &= \langle B\rangle^2 \sinh\phi(\tau),
\end{align}
where the phonon correlation function
\begin{equation}
    \phi(\tau) = \int_{0}^{\infty} \! d\omega\, \frac{J(\omega)}{\omega^2}
    \left[ \coth\!\left( \frac{\beta\hbar\omega}{2} \right)\cos(\omega \tau)
    - i\sin(\omega\tau) \right],
\end{equation}
and thus it follows $G_m(-\tau) = G_m^*(\tau)$. The thermal average of the phonon displacement is
\begin{equation}
    \langle B \rangle = \exp\!\left[ -\frac12 \int_{0}^{\infty} \! d\omega\, 
    \frac{J(\omega)}{\omega^2} \coth\!\left( \frac{\beta \omega}{2} \right) \right],
\end{equation}
with $J(\omega)$ the phonon spectral density, which is given by $J(\omega)=\sum_q\lambda^2_q\delta(\omega-\omega_q)$, and $\beta=\frac{1}{k_B T}$ is inverse temperature. In continuous expression, the phonon spectral density is given by $J(\omega)=\alpha_p\omega^3\exp(-\omega^2/2\omega_b^2)$, where $\alpha_p$ is the exciton-phonon coupling strength and $\omega_b$ is the cutoff frequency of the phonon reservior.

Physically, the prefactor $\langle X\rangle^2$ reflects the reduction of the effective TLS-cavity coupling due to the phonon cloud dressing, which becomes more pronounced at higher temperatures or for stronger coupling to the reservior. The $\cosh\phi$ and $\sinh\phi$ structure indicates that multiphonon processes are naturally included within the polaron framework.

The time evolution of the Hamiltonian $\tilde{H}_I$ is 
\begin{equation}
\tilde{H}_I^{\text{int}}(t) =  \mathcal{T}e^{i(\int_0^t \tilde{H}_S(t')dt' + \tilde{H}_E t)} \tilde{H}_I e^{-i(\int_0^t \tilde{H}_S(t')dt' + \tilde{H}_E t)}.
\end{equation} 
Then the NZ markovian equation in interaction pircture is given by 
\begin{equation}
\frac{\partial \tilde{\rho}^{\text{int}}(t)}{\partial t} = -\int_0^t dt'\, \mathrm{Tr}_B \left\{ [\tilde{H}_I(t), [\tilde{H}_I(t'), \tilde{\rho}^{\text{int}}(t') \rho_B ]] \right\}.
\end{equation}
Tracing out the phonon reservoir and transforming back to the system picture yields
\begin{equation}\label{eq:NZEqn_pn}
    \frac{\partial \tilde{\rho}(t)}{\partial t} = -i[\tilde{H}_S(t), \tilde{\rho}(t)] - \int_0^t R(t,t') dt',
\end{equation}
where $R(t,t')$ is the residual system–reservoir interaction term, which is given by
\begin{equation}\label{eq:NZ_Rpn}
\begin{aligned}
R(t.t')= 
&\sum_{u \in \{g,e\}} G_u(t-t')[X_u, U_S(t,t')X_u\tilde\rho(t')U_S(t',t)]\\
+& \text{H.c.},
\end{aligned}
\end{equation}
and the system propagator $U_S(t,t')$ is defined as
\begin{equation}
    U_S(t,t') = \mathcal{T}\exp \left[ -i \int_{t'}^t d\tau \tilde{H}_S(\tau) \right].
\end{equation}

\section{Derivations of the Dephasing Rate}
The phonon-induced dephasing is characterized in the second order non-Markovian term Eq.~\eqref{eq:NZ_Rpn}, which illustrates the system-phonon reservoir interaction. The phonon-induced Lindblad term can be expanded as
\begin{equation}
\begin{aligned}\label{eq:NZ_R}
R(t,t') = \sum_{u \in \{g,e\}}
& X_u U_S(t,t') X_u \tilde{\rho}(t') U^\dagger_S(t,t') G_u(t-t') \\
&- U_S(t,t') X_u \tilde{\rho}(t') U^\dagger_S(t,t') X_u G_u(t-t') \\
&+ U_S(t,t') \tilde{\rho}(t') X_u U^\dagger_S(t,t') X_u G^*_u(t-t') \\
&- X_u U_S(t,t') \tilde{\rho}(t') X_u U^\dagger_S(t,t') G^*_u(t-t').
\end{aligned}
\end{equation}

The definition $X_u$ follows Eq.~\eqref{eq:Xg} and Eq.~\eqref{eq:Xe}, $U_S(t,t')$ is the propagator operator for $\tilde{H}_S(t)$ in Eq.~\eqref{eq:tildeHS}.

To extract the decay of coherence, we expand the kernel in the truncated single-excitation basis,
\begin{equation}
    R(t,t')=\sum_{m,n,m',n'\in\{1,2,3\}}R^{m'n'}_{mn}(t,t')\tilde\rho_{m'n'}(t')|m\rangle\langle n|,
\end{equation}
Each coefficient $R^{m'n'}_{mn}$ is obtained by projecting Eq.~\eqref{eq:NZ_R} onto $|m\rangle\langle n|$, inserting the matrix elements of $X_u$, $U_S$, and the phonon correlation functions $G_u(t-t')$.

The phonon-induced dephasing rate is obtained from the diagonal coefficient acting on the coherence element $\tilde\rho_{21}$, namely $R^{21}_{21}(t,t')$, which directly governs the time evolution of the off-diagonal density matrix component. Using the unitarity of the system propagator $U_S(t,t')$, one has $|u_{12}|^2 = |u_{21}|^2$ and $|u_{22}|^2 = 1 - |u_{21}|^2$. Substituting these relations into the matrix-element expansion of the kernel yields the dephasing contribution given in Eq.~\eqref{eq:deph_pn} of the main text.

\section{Analytical Solution for the Off-Resonant (Large-Detuning) Rabi Oscillation Regime}

When the interaction between TLS and cavity is suppressed by introducing a large detuning $\Delta$, the system evolution can be described by the following analytic solution. The dynamics of the amplitudes $c_a(t)$ and $c_e(t)$, corresponding to the states $|1, g\rangle$ (one photon in the cavity, the ground state of TLS) and $|0, e\rangle$ (no photon in the cavity, the excited state of TLS), respectively, are given by
\begin{align}
c_a(t) &= \left[ \cos\left( \frac{\tilde{\Delta} t}{2} \right) - i \frac{\Delta}{\tilde{\Delta}} \sin\left( \frac{\tilde{\Delta} t}{2} \right) \right] e^{-i \frac{\Delta}{2} t} \, c_a(0) \\
&\quad - i \frac{2g}{\tilde{\Delta}} \sin\left( \frac{\tilde{\Delta} t}{2} \right) e^{-i \frac{\Delta}{2} t} \, c_e(0), \\
c_e(t) &= \left[ \cos\left( \frac{\tilde{\Delta} t}{2} \right) + i \frac{\Delta}{\tilde{\Delta}} \sin\left( \frac{\tilde{\Delta} t}{2} \right) \right] e^{-i \frac{\Delta}{2} t} \, c_e(0) \\
&\quad - i \frac{2g}{\tilde{\Delta}} \sin\left( \frac{\tilde{\Delta} t}{2} \right) e^{-i \frac{\Delta}{2} t} \, c_a(0),
\end{align}
where
\begin{equation}
\tilde{\Delta} = \sqrt{\Delta^2 + 4g^2}.
\end{equation}

In the limit of large detuning ($|\Delta| \gg g$), the modulus of the amplitudes $c_a(t)$ and $c_e(t)$ remains nearly unchanged, indicating that the population is effectively frozen and negligible transfer occurs between $|1, g\rangle$ and $|0, e\rangle$. However, a time-dependent phase difference between the two components accumulates linearly with time. Specifically, for an initial superposition state, the state vector evolves as
\begin{equation}
|\psi(t)\rangle \approx c_a(0) e^{-i\Delta t} |1, g\rangle + c_e(0) |0, e\rangle,
\end{equation}
demonstrating that the relative phase between the two basis states is not preserved during the detuned period. This phase accumulation becomes significant for quantum control protocols, as it leads to a mismatch between the initial and final phases once the interaction is reactivated.

\bibliography{interaction}

\end{document}